# Pseudo-Hydrogen Passivation: A Novel Way to Calculate Absolute Surface Energy of Zinc Blende (111)/($\bar{1}\bar{1}\bar{1}$) Surface

Yiou Zhang, Jingzhao Zhang, Kinfai Tse, Chunkai Chan, Bei Deng and Junyi Zhu

*Department of Physics, the Chinese University of Hong Kong, Hong Kong*

Abstract

Determining accurate absolute surface energies for polar surfaces of semiconductors has been a great challenge in decades. Here, we propose pseudo-hydrogen passivation to calculate them, using density functional theory approaches. By calculating the energy contribution from pseudo-hydrogen using either a pseudo molecule method or a tetrahedral cluster method, we obtained (111)/($\bar{1}\bar{1}\bar{1}$) surfaces energies of Si, GaP, and ZnS with high self-consistency. This method quantitatively confirms that surface energy is determined by the number and the energy of dangling bonds of surface atoms. Our findings may greatly enhance the basic understandings of different surfaces and lead to novel strategies in the crystal growth.

Absolute surface energies are fundamental physical quantities of solid surfaces with broad implications [1-5]. Equilibrium shapes and morphologies [4,5], equilibrium growth rates [6-8], as well as device performance of semiconductors [9], are directly related to them. The wetting conditions of thin films or super lattices are also determined by these quantities [10]. Novel controlling strategies of growth modes (e.g. by strain or surfactants) are also often determined by them [10-18]. Therefore, determining accurate absolute surface energy is essential in understanding growth processes and in improving the performance of thin-film devices [3,19].

Absolute surface energies of symmetric non-polar surfaces can be calculated with a standard slab method [3]. However, for asymmetric polar surfaces such as zinc blende (111) and ($\bar{1}\bar{1}\bar{1}$) surfaces, it is extremely difficult to separate the anion and cation terminated surfaces, due to the asymmetric nature of slabs [3,19]. Although surface reconstructions and surface kinetic processes can be calculated by passivating the conjugate surfaces with pseudo-H atoms, the absolute surface energy can't be obtained unless the energy of the passivated surface is known [3,19]. Despite the standard treatment of pseudo H passivation in surface calculations, a detailed analysis of the bonding between the surface atom and the pseudo H is lacking and may serve as a key to solve the difficult absolute surface energy problem. One early approach to calculate the absolute surface energies of zinc blende (111) surfaces is to construct a wedge structure and then calculate one polar surface without involving its conjugate surfaces [3]. Based on surface energies calculated by this approach, a common dangling bond rule was also proposed, which states that energies of surface atoms with similar electronic environment are the same regardless of the different surface orientations [3]. Following this method, absolute surface energies of quite a few compound semiconductors along different orientations are calculated [3,19-24]. However, there are a few problems associated with this approach:(1) the surface energies may not be well defined near the edge or corner of the wedge, especially when the wedge size is small, therefore, the size of the wedge structure has to be quite large to reduce edge effect [20], which makes this method rather expensive [19,21]; (2) Pseudo-H near the edge may not be stable [19], which may also affect the accuracy of the calculation. As a result, the calculated absolute surface energies have large errors up to $20 \text{meV}/\text{Å}^2$ [19].

To overcome these problems, we propose a novel method to calculate the absolute surface energies of these surfaces using a pseudo-H passivation approach. Pseudo-H atoms are usually used in zinc blende slab calculations to passivate the dangling bonds of the bottom surface atoms. The pseudo-H atoms carry fractional charge to maintain charge neutrality on the bottom surface, and also stabilize the bottom surface by satisfying electron counting rule (ECR) [25-27]. This passivation ensures that states at the bottom surface are localized and have no interactions with top surface. The energy of the top surface can be directly calculated if the pseudo-H passivation energy can be evaluated. Therefore, a natural and intuitive way to calculate the absolute surface energy is to analyze the pseudo-H passivation process. We show that the energy of the passivated surface can be directly calculated from the pseudo chemical potentials of the pseudo-H atoms attached on the bottom surface. Further, our calculations show that simple pseudo-molecules already give reasonably accurate values of the pseudo chemical potentials. Surface energy calculated from this approach shows comparable self-consistency with the wedge structure calculation, while the computation is much simpler. For high accuracy calculations, we construct a tetrahedral cluster with four equivalent (111)/($\bar{1}\bar{1}\bar{1}$) facets to calculate the pseudo-H chemical potentials and the surface energies show improved self-consistency.

Consider a slab of a binary AB compound of zinc blende structure along [111] direction. The bottom ($\bar{1}\bar{1}\bar{1}$) surface with B-termination is passivated with pseudo-H atom carrying fractional charge denote as $H_B$. The absolute surface energy per unit area of the top (111) surface is then given by

$$\sigma_{top} = \frac{1}{\alpha_{111}}[E_{slab} - n_A\mu_A - n_B\mu_B - n_{H_B}\mu_{H_B} - \alpha_{111}\sigma_{bot}^{pass}], \quad (1)$$

where $E_{slab}$ is the total energy of the slab with bottom surface passivated, $n_A(n_B)$ is the number of A(B) atoms in the slab, $\mu_A(\mu_B)$ is the chemical potential of A(B) atom, $\mu_{H_B}$ is the chemical potential of pseudo-H $H_B$, $\alpha_{111}$ is the area of (111) surface and $\sigma_{bot}^{pass}$ is the surface energy of the passivated bottom surface. Assuming a thermodynamic equilibrium between the bulk and surface, we can write

$$\mu_A + \mu_B = E_{AB} = E_A + E_B + \Delta H_f(AB), \quad (2)$$

where $E_{AB}$, $E_A$ and $E_B$ are total energy of corresponding bulk solid, and $\Delta H_f(AB)$ is the formation enthalpy of AB compound. To avoid presence of either solid A or solid B, it is required that $\Delta\mu_A = \mu_A - E_A$ satisfy

$$\Delta H_f(AB) \leq \Delta\mu_A \leq 0, \quad (3)$$

which the limits correspond to the A-poor and A-rich limit. On the right hand side (RHS) of Eq. (1), all terms can be easily determined from first-principle calculations except for $\mu_{H_B}$ and $\sigma_{bot}^{pass}$, which is the major focus of this Letter.

To calculate these terms, we define a pseudo chemical potential $\hat{\mu}_{H_B}$ for $H_B$ by considering the sum of two terms, so that

$$n_{H_B}\hat{\mu}_{H_B} = \alpha_{111}\sigma_{bot}^{pass} + n_{H_B}\mu_{H_B}, \quad (4)$$

and Eq. (1) can be rewritten as

$$\sigma_{top} = \frac{1}{\alpha_{111}}[E_{slab} - n_A\mu_A - n_B\mu_B - n_{H_B}\hat{\mu}_{H_B}]. \quad (5)$$

The pseudo chemical potential describes the energy gain from adding one pseudo-H atom and passivating one dangling bond on the bottom surface with this pseudo-H atom. This pseudo chemical potential can be decomposed into

$$\hat{\mu}_{H_B} = \mu_{H_B} + [\delta E_{int} + \delta E_{env}], (6)$$

where the former part is the chemical potential of $H_B$ atom, and the latter part in bracket is the binding energy between the surface atom and the pseudo-H atom. This binding energy is just the energy of the passivated surface, divided by the number of passivated bond rather than surface area. It can be further decomposed into $\delta E_{int}$ due to the intrinsic property of the surface atom, and $\delta E_{env}$ due to electronic environment. Since passivated surfaces satisfy charge neutrality and ECR [25-27], contribution from the environment is expected to be localized, and the major contribution comes from the local electronic environment around the pseudo-H atoms. It is difficult to calculate each individual part of the pseudo chemical potential, but the summation of all parts can be estimated under a local electronic environment similar to that of the surface atoms. *This transforms the problem of calculating the energy of individual polar surface to a problem of estimating energy of bonds between surface atoms and pseudo-H atoms with a similar electronic environment.* Such estimation only requires reproducing a local electronic environment similar to that of the surface atoms and pseudo-H atoms on the surface, but not the overall structure and symmetry of the surface. Therefore, this method is generally applicable to any crystal planes, as long as we can determine the pseudo chemical potential of $H_B$ with the similar local environment on the surface. Also, if we passivate the top surface with pseudo-H atom $H_A$, left hand side (LHS) of Eq. (5) becomes $\sigma_{top}^{pass} = n_{H_A}\hat{\mu}_{H_A}/\alpha_{111}$, and we obtain

$$n_{H_A}\hat{\mu}_{H_A} + n_{H_B}\hat{\mu}_{H_B} = E'_{slab} - n_A\mu_A - n_B\mu_B, \quad (7)$$

where $E'_{slab}$ denotes a total energy of the slab with both surfaces passivated. Eq. (7) describes the energy of all bonds between surface atoms and pseudo-H atoms, whereas LHS is calculated from pseudo chemical potentials and RHS is from standard slab calculation. Therefore, Eq. (7) can be used to directly determine the difference between the obtained pseudo chemical potentials and the standard slab calculations, which defines the self-consistency of the calculation.

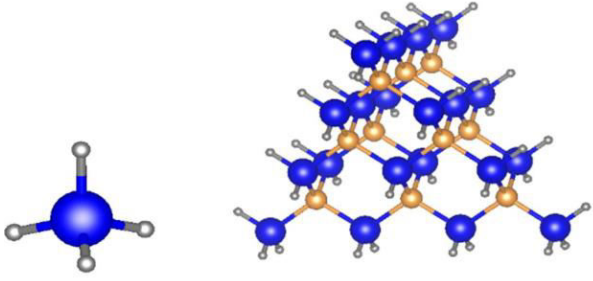

Fig. 1. (Color online) Schematic illustration of the structure of a pseudo-molecule and that of a tetrahedral cluster. The largest blue balls represent A atom, the moderate orange balls represent B atom, and the smallest balls represent pseudo-H atom $H_A$. For tetrahedral cluster in (b), the number of A atoms on the edge is n = 4.

Here, we propose two ways to determine the pseudo chemical potential for the $(111)/(\bar{1}\bar{1}\bar{1})$ surface, one from a simple pseudo-molecule and the other from a tetrahedral cluster. For the pseudo-molecule method, we construct a $CH_4$-like molecule, with A(B) atom at the center of a tetrahedron bonded to four $H_A(H_B)$ atom at the corner of the tetrahedron, as shown in Fig. 1(a). It can be viewed as passivating four dangling bonds of a free-standing atom by pseudo-H. Since there are four bonds between center atom and pseudo-H, we can determine the pseudo chemical potential by

$$\hat{\mu}_{H_A} = (E_{molecule} - \mu_A)/4. \quad (8)$$

Using this method, the chemical potential of $H_A$ and intrinsic contribution to the binding energy can be calculated, but it does not reproduce the local electronic environment. This method is straightforward and computationally inexpensive, nevertheless yields a fairly accurate result. Thus, it can be taken as the $0^{th}$ order approximation for the pseudo chemical potential of the pseudo-H.

The cluster method, in addition, reproduces local electronic environment similar to that on the surface. The structure is shown in Fig. 1(b). The cluster contains four (111) facets and all the dangling bonds on the surface are passivated by the corresponding pseudo-H atoms. The size of the cluster can be identified by $n$, the number of atoms on the edge. From the figure, we can identify surface atoms with different local environment. For each A atom on the corner, it is bonded to one B atom and three $H_A$ atoms; for each A atom on the edge but not on the corner, it is bonded to two B atoms and two $H_A$ atoms; for each A atom on the face of the tetrahedron, it is bonded to three B atoms and one $H_A$ atom (similar to (111) surface). We can denote the pseudo chemical potentials under these three conditions as $\hat{\mu}_{H_A}^{cor}$, $\hat{\mu}_{H_A}^{edge}$ and $\hat{\mu}_{H_A}^{face}$ respectively. Local electronic environment of $H_A$ atoms on the face of the clusters is similar to that of $H_A$ atoms on (111) surface. Therefore, $\hat{\mu}_{H_A}^{face}$ is a good approximation to $\hat{\mu}_{H_A}$ on (111) surface. Since number of A atoms, B atoms and pseudo-H atoms can all be expressed by the cluster size n, we write the total energy of the cluster as

$$E_{cluster}(n) = \frac{1}{6}n(n+1)(n+2)\mu_A + \frac{1}{6}n(n-1)(n+1)(E_{AB} - \mu_A) + 2(n-2)(n-3)\,\hat{\mu}_{H_A}^{face} + 12(n-2)\,\hat{\mu}_{H_A}^{edge} + 12\,\hat{\mu}_{H_A}^{cor},$$

(9)

where $E_{AB}$ here is explicitly taken as a variable allowing for small deviations from the bulk energy. Such a constant shift of bulk atom energy has previously been observed from standard slab calculations [28]. By calculating four clusters of different size, we can solve the Eq. (7) and obtain $\hat{\mu}_{H_A}^{face}$ as a good approximation to $\hat{\mu}_{H_A}$ on (111) surface. Similarly, we can determine $\hat{\mu}_{H_B}$

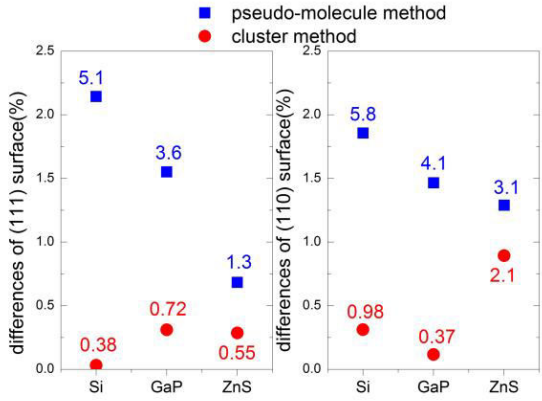

Fig. 2. A summary of results from pseudo-molecule method and tetrahedral cluster method. Both percentage differences and differences per surface area are calculated and listed. Vertical scale shows percentage differences between the obtained pseudo chemical potentials and slab calculations. Energy differences per surface area (meV/Å$^2$) are labeled on the figure for each point.

on ($\bar{1}\bar{1}\bar{1}$) surface by swapping B with A, and $H_A$ with $H_B$.

Most of the calculations were done with Generalized Gradient Approximation (GGA). As GGA functional usually gives a smaller band-gap than experimental value, which affects the accuracy of the surface energies [19], we also performed calculations with screened hybrid functional on slabs and pseudo-molecules of GaP to test the difference between GGA functional and hybrid functional. Results indicate that our proposed method is general and is not functional dependent. Additional details of the calculations are listed in the Supplemental Materials.

For a proof of principle, we considered three semiconductors, Si, GaP and ZnS. Only the absolute energy of Si surfaces can be calculated by constructing symmetric slabs. However, for compound semiconductors, we can construct slabs with both surfaces passivated, and calculate the energy of the fully passivated slabs, where two different kinds of pseudo-H atoms are involved. Then by making use of Eq. (7), we can obtain the sum of the pseudo chemical potentials with the standard slab calculations. The differences per surface area between the sum based on the cluster (or pseudo molecule) method and the slab method were used to check the self-consistency of our method as well as to estimate the errors of the obtained surface energies. The percentage differences were also calculated. Throughout the calculation, the chemical potentials of Ga and that of Zn are taken at the rich limit for GaP and ZnS, respectively.

For slab calculations, we considered the polar 111/($\bar{1}\bar{1}\bar{1}$) surfaces, and the non-polar (110) surfaces. Surface energy of (110) surface can be determined from the standard slab calculation, so this provides additional validity check with our method. Also, each surface atom on unreconstructed (110) surface contains one dangling bond, similar to (111)/($\bar{1}\bar{1}\bar{1}$) surface.

After the slab calculations, we calculated pseudo chemical potential from pseudo-molecule method. Results are summarized in Fig.2. For 111/($\bar{1}\bar{1}\bar{1}$) surface, difference between the slab calculations and the pseudo chemical potentials calculations are all within $6 meV/Å^2$. Hybrid functional calculations of GaP show a difference of $9.1 meV/Å^2$, slightly larger than that for GGA calculations. Calculations on (110) surfaces of Si, GaP and ZnS also show differences within $6 meV/Å^2$. Previous calculations based on wedge structure have $3 meV/Å^2$[3] and $20 meV/Å^2$[19] differences for GaAs and zinc blend GaN respectively. Therefore, these results show good accuracies comparable to the wedge structure calculation, whereas the calculations are much simpler.

For the tetrahedral cluster method, we construct clusters with different sizes from n=2 to n=9. Different selections of clusters yield different accuracies of the results, as explained in the Supplemental Materials. Converged results from these pseudo chemical potentials are summarized in Fig. 2. For both (111)/($\bar{1}\bar{1}\bar{1}$) and (110) surfaces of Si, GaP and ZnS, the differences are within $1 meV/Å^2$ with slab calculations, except for ZnS (110) surface. These results show large improvement of accuracy over the wedge structure calculations. Even though we construct the clusters based on (111)/($\bar{1}\bar{1}\bar{1}$) facets, it also works very well for (110) surface, which confirms that number of pseudo-H atoms attached to each surface atom should be the major contribution to the accurate pseudo chemical potential of pseudo-H. In another word, surface energy is directly determined by the dangling bonds of surface atoms.

Table I. Calculated absolute surface energies of unreconstructed (111), ($\bar{1}\bar{1}\bar{1}$) and (110) surfaces of Si, GaP and ZnS, based on pseudo chemical potentials from cluster method, in units of $meV/Å^2$. Values in parentheses are differences with surface energies calculated from standard slab calculation.

|  | Si | GaP | ZnS |
|---|---|---|---|
| (111) | 97.8(-0.4) | 87.5 | 88.2 |
| ($\bar{1}\bar{1}\bar{1}$) | ----- | 100.8 | 88.6 |
| (110) | 92.2(-1.0) | 46.3(-0.4) | 23.9(-2.1) |

From the estimation of pseudo chemical potential of pseudo-H atoms on (111)/($\bar{1}\bar{1}\bar{1}$) and (110) surfaces, we can see that major contributions of the pseudo chemical potential are from the chemical potential of pseudo-H $\mu_{H_{A/B}}$, and intrinsic contribution to the binding energy $\delta E_{int}$. The contribution from the local electronic environment, $\delta E_{env}$, is not significant. This is because on all the slabs and clusters, pseudo-H atoms have enough space to relax, and the surrounding local electronic environment only serves as a perturbation on the binding energy. Therefore, we have $\delta E_{env} \ll \mu_{H_{A/B}} + \delta E_{int}$. Also for $\delta E_{env}$, the contribution from the 1$^{st}$ nearest neighbors of the surface atoms is much larger than that from the rest. Since the cluster method gives correct 1$^{st}$ nearest neighbors for surface atoms on both 111/($\bar{1}\bar{1}\bar{1}$) and (110) surfaces, both results are very accurate. This shows general applicability of the pseudo chemical potential to determine the absolute surface energy of polar surfaces, regardless of the overall geometry of the surfaces.

Several advantages can be achieved by using our proposed methods. Since pseudo-molecule method takes both the chemical potential of pseudo-H atoms and the intrinsic contributions to the binding energy between the pseudo H and the surface atoms, it has comparable accuracies as the wedge structure calculations. More importantly, this method is much simpler than the wedge method and can be easily applied to other surfaces, especially to polar surfaces other than (111)/($\bar{1}\bar{1}\bar{1}$) surfaces, where wedges may be difficult to construct [21]. The stability issues of the wedge methods can be avoided by the cluster methods, because of the high symmetry of the tetrahedral structures. The atomic structures are allowed to fully relax without constraints in all the cluster calculations. Therefore, the inaccuracy caused by instability of pseudo-H atoms and the finite size effects in wedge structure calculations can be largely avoided. Even though 4 clusters are essential for determining the surface energies while only two wedge structures are needed, two of the clusters are very small and easy to calculate. The large size difference between the large cluster and the small cluster greatly improves the accuracy of the method. Generally speaking, our methods are expected to save computing time and yield high accuracies.

From Table I, we can conclude the general trends of the surface energies for different compounds. For unreconstructed (111)/($\bar{1}\bar{1}\bar{1}$) surfaces, surface energies for Si, GaP and ZnS follow the trend of their cohesive energies, since the electron redistribution is not significant and surface energies are just directly determined by the dangling bonds on the surface. However, surface energies on (110) surfaces decrease sharply with the increase of the iconicity of the materials, because in those compounds, ECR can probably be better satisfied when electrons in cation dangling bonds are transferred to anion dangling bonds due to the large electronegativity difference between them [25-27]. Also, the energy cost from forming the dimer-like structures on the surfaces is smaller for compounds with strong ionicity because the bond strength in such compounds is weaker than that in covalently bonded compounds.

In summary, we have proposed a new method to calculate surface energy of (111)/($\bar{1}\bar{1}\bar{1}$) polar surface of zinc blende structure, based on pseudo-H passivation analysis. Tests on (111)/($\bar{1}\bar{1}\bar{1}$) and (110) surface of Si, GaP and ZnS show very accurate results and good consistency with slab calculations. This method is not restricted to (111)/($\bar{1}\bar{1}\bar{1}$) surfaces, and it is generally applicable to other surfaces of many other types of crystals. The 0$^{th}$ order approximation of the method yields reasonable accuracy that is comparable with wedge methods, but saves much computing time. The high order approach largely improves the accuracy of the absolute surface energy calculations, which are expected to provide very important physical insights in crystal growth techniques, thin film properties controls, and device performance enhancement. In particular, this method can give accurate surface energies of c/-c planes of wurtzite structures [29]. Our method also quantitatively confirms that surface energy is directly determined by the number and energy of dangling bonds of surface atoms for the first time.

We would like to thank Su-huai Wei for helpful discussions. Computing resources were provided by the High Performance Cluster Computing Centre, Hong Kong Baptist University. This work was supported by the start-up funding and direct grant with the Project code of 4053134 at CUHK.

# Supplemental material: Pseudo-Hydrogen Passivation: A Novel Way to Calculate Absolute Surface Energy of Zinc Blende (111)/($\overline{1}\overline{1}\overline{1}$) Surface

Yiou Zhang, Jingzhao Zhang, Kinfai Tse, Chunkai Chan, Bei Deng and Junyi Zhu
*Department of Physics, the Chinese University of Hong Kong, Hong Kong*

The total energy calculations of bulks, slabs and clusters were based on Density Functional Theory [1,2] as implemented in VASP code [3,4], with a plane wave basis set [5,6]. The energy cutoff of the plane wave was set at 400eV. PBE Generalized Gradient Approximation (GGA) functional [7] was used for GGA functional calculations. Screened hybrid functional of Heyd, Scuseria, and Ernzerhof (HSE) [8,9] was used for hybrid functional calculations.

All the slab calculations were performed on (1×1) slabs, with (10×10×1) Monkhorst-Pack [10] k-point mesh for integration over Brillouin zone for GGA calculations and (4×4×1) for hybrid functional calculations. The slabs and clusters were separated by at least 15Å vacuum. Pseudo-H atoms with charge q=0.5e, 0.75e, 1.25e, and 1.5e were used to passivate dangling bonds of S, P, Ga, and Zn atoms, respectively. For Si, the passivation is done by true H. All the atoms in the slab and cluster were allowed to relax until forces converged to less than 0.005eV/Å.

Slabs along [111] direction contain 9 bi-layers, with both surfaces passivated by the corresponding pseudo-H atoms. Slabs along [110] direction contain 12 layers, and calculations were done both for slabs with both surfaces un-passivated and slabs with one surface passivated. Convergence tests are performed by increasing the number of layers in the slabs, and the results indicate that the obtained numerical errors of the surface energies are less than 0.5meV/Å$^2$..

Calculations on conventional unit cells of Si, GaP, ZnS yield good consistence with experimental results. The calculated lattice constants are 5.47Å (experiment [11]: 5.43Å), 5.50Å (experiment [12]: 5.45Å) and 5.44Å (experiment [13]: 5.41Å) for Si, GaP and ZnS respectively, which all show differences within 1%.

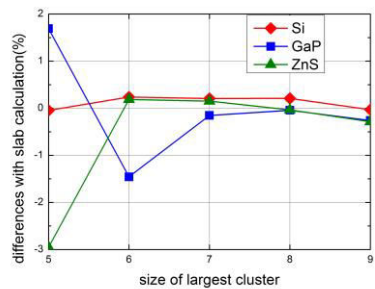

Fig.S1. Differences between slab calculations of 111/($\overline{1}\overline{1}\overline{1}$) surfaces and pseudo chemical potentials obtained from different selections of clusters. The smallest clusters with n=2 are always included and horizontal axis denotes the size of the largest cluster included. The difference between the Si (111) surface energy based on pseudo chemical potential calculation and that based on slab calculation is also included as a reference. Slight increase of differences at n=9 for GaP and ZnS are mainly due to numerical errors between slab calculations and cluster calculations.

For the tetrahedral cluster method, although any four clusters can be used to solve for Eq. (8), different selections in fact give different results, as shown in Fig. S1. To make fair comparison between different systems, the percentage differences rather than differences per surface area were used. The variation of the results from clusters of different sizes is because we determine the pseudo chemical potentials from the energy differences between clusters. If energy differences between chosen clusters are large, the errors in the total energies of those clusters will be less significant. Hence the obtained pseudo chemical potentials will be more accurate. Therefore, in all the calculations, two smallest clusters and two largest clusters are chosen in the linear equation set to improve the accuracy. As can be seen from Fig. S1, the last three points show good convergence, with percentage differences less than 0.3% (1meV/Å$^2$ in term of energy difference per surface area), and the obtained $E_{tot}$(AB) calculated from Eq. (8) also shows only a few meV difference with that from bulk calculations. Therefore, we can take theses converged results as the pseudo chemical potentials obtained by cluster method, and the remaining differences as the errors of our method.